\documentstyle[sprocl]{article}

\input{psfig}

\def\etal{{et al.~}}
\def\kms{\>{\rm km}\,{\rm s}^{-1}}
\def\Msun{\>{\rm M_{\odot}}}
\def\pc{\>{\rm pc}}
\def\kpc{\>{\rm kpc}}

\begin{document}

\title{The Shapes of Galaxies and Their Halos as Traced by Stars:\\
       The Milky Way Dark Halo and The LMC Disk}

\author{Roeland P.~van der Marel}

\address{STScI, 3700 San Martin Drive,
Baltimore, MD 21218, USA\\Email: marel@stsci.edu}


\maketitle

\abstracts{Stars and their kinematics provide one of the tools
available for studies of the shapes of galaxies and their halos. In
this review I focus on two specific applications: the shape of the
Milky Way dark halo and the shape of the LMC disk. The former is
constrained by a variety of observations, but an accurate
determination of the axial ratio $q_{\rm DH}$ remains elusive. A very
flattened Milky Way dark halo with $q_{\rm DH} \leq 0.4$ is ruled out,
and values $q_{\rm DH} \geq 0.7$ appear most consistent with the
data. Near-IR surveys have revealed that the LMC disk is not
approximately circular, as long believed, but instead has an axial
ratio of $0.7$ in the disk plane. The elongation is perpendicular to
the Magellanic Stream, indicating that it is most likely due to the
tidal force of the Milky Way. Equilibrium dynamical modeling of
galaxies is important for many applications. At the same time,
detailed studies of tidal effects and tidal streams have the potential
to improve our understanding of both the Milky Way dark halo and the
structure of satellite galaxies such as the LMC.}

\section{Introduction}

A large variety of tracers can be used to study the three-dimensional
shapes of galaxies and their dark halos. Other contributions to this
volume discuss the use of cold gas (e.g., HI), hot gas (X-rays),
globular clusters, planetary nebulae, satellite galaxies, and
gravitational lensing (both weak and strong) to study this important
topic. By contrast, the present review focuses on the constraints that
can be obtained from studies of stars and their kinematics. It is not
possible in the context of this relatively short paper to review this
topic fully for all possible classes of galaxies. So instead I focus
here on two subjects for which studies of stars and their kinematics
are particularly relevant: the three-dimensional shape of the Milky
Way dark halo, and the intrinsic shape of the Large Magellanic Cloud.

\section{The Shape of the Milky Way Dark Halo}
\label{s:MW}

The Milky Way consists of various separate components, including the
thin disk, thick disk, central bulge, central black hole, metal-poor
halo and dark halo. The general structure of the Milky Way and its
various components has been reviewed by many previous authors,
including, e.g., Freeman (1987), Gilmore, Wyse \& Kuijken (1989),
Gilmore, King \& van der Kruit (1990), Majewski (1993) and Binney \&
Merrifield (1998). Here I focus on the shape of the Galactic Dark
Halo, a subject previously reviewed by Sackett (1998).

\subsection{Constraints from Metal-Poor Halo Star Kinematics}
\label{ss:MWhalo}

The density distribution of the metal-poor halo has been studied using
blue horizontal branch stars, RR Lyrae stars, stars counts and
globular clusters. These studies have shown that the metal-poor halo
is approximately a spheroidal system with an axial ratio $q_{\rm MWH}$
and a mass density profile that falls radially as a power-law, $\rho
\propto r^{-n}$. An unweighted average of the many determinations
available in the literature (as cited in, e.g., Chen \etal 2001)
yields $q_{\rm MWH} = 0.73 \pm 0.11$ and $n = 2.95 \pm 0.37$ (where
the quoted errors are the RMS variations between different
studies). In the solar neighborhood one can determine the
three-dimensional stellar velocity ellipsoid of the metal-weak halo
stars, because both line-of-sight velocities and proper motions can be
measured. An unweighted average of determinations in the literature
(as cited in, e.g., Martin \& Morrison 1998) yields
$(\sigma_R,\sigma_{\phi},\sigma_z) = (155 \pm 16, 105 \pm 7, 97 \pm 7)
\kms$. One can show that these observations yield a lower limit on the
axial ratio $q_{\rm DH}$ of the Milky Way dark halo.  Hydrostatic
equilibrium in an oblate collisionless system requires that there is
more pressure parallel to the equatorial plane than perpendicular to
it, i.e., ${\cal R} \equiv (\sigma_R^2 + \sigma_{\phi}^2) / 2
\sigma_z^2 > 1$. The flatter the Milky Way dark halo, the smaller the
ratio ${\cal R}$ for the metal-weak halo stars must be to sustain a
system of axial ratio $q_{\rm MWH}$.  An analysis based on the Jeans
equations as in van der Marel (1991) yields the strict limit that
$q_{\rm DH} > 0.4$, with all larger values allowed (depending on the
unknown details of the stellar phase-space distribution
function). This rules out certain models for galactic dark halos, such
as those in which the dark matter all resides in the disk (e.g.,
Pfenniger, Combes \& Martinet 1994).

\subsection{Constraints from Disk Star Kinematics}
\label{ss:MWdisk}

Observations of the kinematics of stars in the Milky Way disk have
been used to estimate that the total surface mass density of material
within $1.1 \kpc$ from the equatorial plane is $\Sigma_{\rm tot} = 71
\pm 6 \Msun/\pc^2$ (Kuijken \& Gilmore 1991). Approximately half of
this mass can be accounted for by the known luminous components of the
Milky Way. From star count analyses Kuijken \& Gilmore (1989) find
that $\Sigma_{\rm lum} = 35 \pm 5 \Msun/\pc^2$, while Gould, Bahcall
\& Flynn find $\Sigma_{\rm lum} = 26 \pm 4 \Msun/\pc^2$. For any
family of dark halo models that reproduce the Milky Way rotation
curve, the amount of dark matter $\Sigma_{\rm dark}$ within $1.1 \kpc$
from the equatorial plane increases as $q_{\rm DH}$ is
decreased. Given that $\Sigma_{\rm tot} = \Sigma_{\rm lum} +
\Sigma_{\rm dark}$, this yields a constraint on $q_{\rm DH}$ (Olling
\& Merrifield 2000). However, this constraint is not much more
stringent than $q_{\rm DH} = 0.7 \pm 0.3$, given the uncertainties in
$\Sigma_{\rm lum}$ and the Galactic constants $R_0$ (the solar
distance to the Galactic center) and $V_0$ (the circular velocity at
the solar radius).

Other kinematical properties of disk stars provide the only useful
constraints on the in-plane ellipticity $(b/a)_{\rm DH}$ of the Milky
Way dark halo. Evidence for non-circularity derives primarily from the
local value of $\sigma_{\phi} / \sigma_R$ for thin disk stars, and
from discrepancies between the Galactic rotation curve at $R > R_0$
when estimated using stellar and gaseous tracers respectively. Kuijken
\& Tremaine (1994) conclude tentatively from this evidence that the
gravitational potential has an ellipticity of $\sim 0.08$, with the
sun positioned roughly along the minor axis. The gravitational
potential of a mass distribution is always rounder than the density
distribution, and the in-plane ellipticity of the dark halo density
must therefore be $(b/a)_{\rm DH} \approx 0.2$. This would be somewhat
surprising, given that studies of other disk galaxies have found that
their ellipticities do not typically exceed $\sim 0.1$ (Franx \& de
Zeeuw 1992; Rix \& Zaritsky 1995; Schoenmakers, Franx \& de Zeeuw
1997).

\subsection{Constraints from Tidal Streams of Stars in the Halo}
\label{ss:MWstreams}

When an infalling satellite is tidally disrupted by a galaxy, the
tidally stripped material will phase-mix along the satellite
orbit. This gives rise to tidal streamers (Johnston, Hernquist \&
Bolte 1996). It has been suggested that much of the metal-weak halo
was built up in this way (Searle \& Zinn 1978), and evidence for this
is now accumulating (Helmi \etal 1999). In a circular potential the
tidal stream will move in a planar orbit, and will appear to an
observer approximately as a great circle on the sky. By contrast, in
an axisymmetric potential the orbit will precess, and the projected
distribution on the sky will be considerably more spread out. Using a
study of carbon stars, Ibata \etal (2001) recently found a stream that
is associated with the Sagittarius dwarf galaxy. They argue that the
stream has only a small width on the sky, and from numerical
simulations it is inferred that this implies $q_{\rm DH} > 0.7$ for
the Milky Way dark halo. However, the stream has so far been
delineated by only 38 carbon stars. So while this is a promising new
method for studying the Milky Way dark halo shape, the resulting
constraint on $q_{\rm DH}$ should be considered somewhat tentative at
present, given the limited statistics.

\subsection{Constraints from Other Tracers}
\label{ss:MWother}

Although this review focuses on stars and their kinematics, it is
useful in the present context to discuss what other tracers and
methods can be used to constrain the shape of the Milky Way dark
halo.\\

\noindent {\tt [HI flaring]} The thickness of the HI layer of the
Milky Way increases with galactocentric distance. The exact amount of
disk flaring is directly related to the axial ratio of the dark halo
through the equations of vertical hydrostatic equilibrium, as modeled
in detail by Olling \& Merrifield (2000). Unfortunately, the results
of this analysis depend sensitively on the poorly known Galactic
constants $R_0$ and $V_0$. When acceptable margins on these constants
are taken into account, the observed HI flaring can be fit with models
in which $q_{\rm DH}$ is anywhere between $0.5$ and prolate values
$>1$. This degeneracy can be lessened when the models are required to
also fit the constraints from the stellar kinematics of the thin disk,
as described in Section~\ref{ss:MWdisk} (because the latter have a
different functional dependence on the Galactic constants). Olling \&
Merrifield (2000) advocate $q_{\rm DH} \sim 0.8$ as their best-fit
value, but this is obtained for values $R_0 \approx 7.6 \kpc$ and $V_0
\approx 190 \kms$ that differ considerably from the commonly accepted
values. In the absence of accurate independent determinations of $R_0$
and $V_0$ these results therefore remain tentative.\\

\noindent {\tt [Microlensing]} In a dark halo composed of massive
compact halo objects (MACHOs) the optical depth for microlensing in
different directions will depend on the axial ratio $q_{\rm DH}$ of
the dark halo. Sackett \& Gould (1993) argued that the ratio of the
optical depths towards the LMC and the SMC can thus be used to
estimate $q_{\rm DH}$. Although surveys towards the LMC and the SMC
have indeed detected microlensing, the prospects for a determination
of $q_{\rm DH}$ are not promising. The latest results from the MACHO
collaboration indicate that only $\sim 20$\% of the dark halo is
composed of MACHOs (Alcock \etal 2000). Hence, microlensing can only
yield constraints on $q_{\rm DH}$ if the MACHOs trace the rest of dark
halo material, and this is unclear as long as the nature of the MACHOs
remains unknown. In addition, there remains controversy concerning the
contribution of LMC and SMC self-lensing to the observed optical
depths (e.g., Sahu 1994; Afonso \etal 2000), which complications any
interpretation of the data.\\

\section{The Intrinsic Shape of the LMC}
\label{s:LMC}

The LMC is our nearest significant neighbor galaxy. While for the
Milky Way the shape of the dark halo continues to be one of the
outstanding questions, for the LMC it is still an important question
what the shape of the stellar component itself is. This is of
fundamental importance for several issues, including the study of the
Milky Way dark halo through modeling of the Magellanic Stream and the
tidal disruption of the Magellanic Clouds (e.g., Moore \& Davis 1994;
Lin, Jones  \& Klemola 1995; Gardiner \& Noguchi 1996) and the study of
compact objects in the dark halo through microlensing studies (e.g.,
Sahu 1994; Alcock \etal 2000).

\subsection{The Vertical Structure of the LMC}
\label{ss:LMCvert}

The LMC is believed to be approximately planar. This is supported by:
(a) the small vertical scale height ($< 0.5 \kpc$) indicated by the
line-of-sight velocity dispersion of Long Period Variables (Bessell,
Freeman \& Wood 1986), star clusters (Freeman, Illingworth \& Oemler
1983; Schommer \etal 1992), planetary nebulae (Meatheringham \etal
1988) and carbon-rich AGB stars (Alves \& Nelson 2000); and (b) the
relatively small scatter in the period-luminosity-color relationships
for Cepheids (Caldwell \& Coulson 1986) and Miras (Feast \etal
1989). There is some evidence for an increasing scale height at large
radii (Alves \& Nelson 2000). There is no evidence for a halo
component in the LMC comparable to that of our own Galaxy (Freeman
\etal 1983), although it has been a topic of debate whether the LMC
contains secondary populations that do not reside in the main disk
plane (e.g., Luks \& Rohlfs 1992; Zaritsky \& Lin 1997; Zaritsky \etal
1999; Weinberg \& Nikolaev 2000; Zhao \& Evans 2000).

\subsection{The Viewing Angles of the LMC Disk Plane}
\label{ss:LMCview}

It has only very recently been possible to accurately determine the
viewing angles of the LMC disk plane: the inclination angle $i$ and
the line-of-nodes position angle $\Theta$. The method relies on the
fact that one side of the LMC disk plane is closer to us than the
other, which causes stars on one side of the LMC to be brighter than
those on the opposite side. Van der Marel \& Cioni (2001) detected
this effect using an analysis of spatial variations in the apparent
magnitude of features in the color-magnitude diagrams extracted from
the near-IR surveys DENIS (e.g., Cioni \etal 2000a) and 2MASS (e.g.,
Nikolaev \& Weinberg 2000). Sinusoidal brightness variations with a
peak-to-peak amplitude of $\sim 0.25$ mag were detected as function of
position angle. The same variations are detected for asymptotic giant
branch (AGB) stars (using the mode of their luminosity function) and
for red giant branch (RGB) stars (using the tip of their luminosity
function), and these variations are seen consistently in all of the
near-IR DENIS and 2MASS photometric bands. The best fitting geometric
model of an inclined plane yields an inclination angle $i =
34.7^{\circ} \pm 6.2^{\circ}$ and line-of-nodes position angle $\Theta
= 122.5^{\circ}\pm 8.3^{\circ}$.

\subsection{The Shape of the LMC Disk}
\label{ss:LMCelong}

The LMC morphology has been studied with many different tracers
(reviewed by Westerlund 1997), including e.g.~stellar clusters, HII
regions, supergiants, planetary nebulae, HI emission and non-thermal
radio emission. However, near-IR surveys provide the most accurate
view because of the large statistics and insensitivity to dust
absorption. van der Marel (2001) used the 2MASS and DENIS data to
construct a star count map of RGB and AGB stars. The resulting LMC
image shows the well-known bar, but indicates that the
intermediate-age/old stellar component is otherwise quite smooth (see
also Nikolaev \& Weinberg 2000, Cioni \etal 2000b). This contrasts
with the younger populations that dominate the light in optical
images. Ellipse fitting shows that the position angle and ellipticity
profile have large radial variations at small radii, but converge to
${\rm PA}_{\rm maj} = 189.3^{\circ} \pm 1.4^{\circ}$ and $\epsilon =
0.199 \pm 0.008$ for $r > 5^{\circ}$.

Van der Marel (2001) stressed the importance of the fact that $\Theta$
differs from ${\rm PA}_{\rm maj}$. This indicates that the LMC disk is
not circular. Deprojection of the near-IR starcount map yields an
intrinsic ellipticity $0.31$ in the outer parts of the LMC. This
elongation had not been previously recognized, because traditionally
the line of nodes position angle $\Theta$ had generally been
determined under the (incorrect) assumption that the LMC is circular
(as discussed and reviewed in van der Marel \& Cioni 2001).

\subsection{The Tidal Effect of the Milky Way}
\label{ss:LMCtidal}

The elongation of the LMC is considerably larger than typical for disk
galaxies (see the discussion in Section~\ref{ss:MWdisk}). Weinberg
(2000) recently stressed the importance of the Galactic tidal field
for the structure of the LMC. To lowest order one would expect the
main body of the LMC to become elongated in the direction of the tidal
force, i.e., towards the Galactic Center. By contrast, material that
is tidally stripped will phase mix along the orbit (see
Section~\ref{ss:MWstreams}). For an orbit that is not too far from
being circular, one thus expects the elongation of the main body to be
perpendicular to any tidal streams that emanate from it (a generic
feature of satellite disruption that is often seen in numerical
simulations and observations, e.g., Johnston 1998; Odenkirchen \etal
2001). The results of van der Marel (2001) show indeed that LMC is
elongated in the general direction of the Galactic center, and is
elongated perpendicular to the Magellanic Stream and the velocity
vector of the LMC center of mass. This suggests that the elongation of
the LMC has been induced by the tidal force of the Milky Way.

\section{Conclusions}
\label{s:conc}

Stars and their kinematics provide an important tool for the study of
the shapes of galaxies and their dark halos. The discussions in this
review show that while for the nearest galaxies some important lessons
have been learned, other issues continue to be poorly understood. The
Milky Way dark halo cannot be very flattened, but otherwise its axial
ratio is not well constrained. For the LMC we have only recently
learned that its disk is quite elongated, and that the tidal effect of
the Milky Way on LMC structure may be larger than previously
believed. It appears that detailed studies of tidal effects and tidal
streams have the potential to improve our understanding of both the
Milky Way dark halo and the structure of satellite galaxies such as
the LMC.



\section*{References}

\begin{thebibliography}{99}

\bibitem{Afo00}
Afonso, C., et al. 2000, ApJ, 532, 340

\bibitem{Alc00}
Alcock, C., et al. 2000, ApJ, 542, 281

\bibitem{Alv00}
Alves, D. R., \& Nelson, C. A. 2000, ApJ, 542, 789

\bibitem{Bes86}
Bessell, M. S., Freeman, K. C., \& Wood, P. R. 1986, ApJ, 310, 710

\bibitem{Bin98}
Binney, J. J., \& Merrifield, M. 1998, 
Galactic Astronomy (Princeton: Princeton University Press)

\bibitem{Cal86}
Caldwell, J. A. R., \& Coulson, I. M. 1986, MNRAS, 218, 223

\bibitem{Che01}
Chen, B., et al. 2001, ApJ, 553, 184

\bibitem{Cio00a}
Cioni, M. R., et al. 2000a, A\&AS, 144, 235

\bibitem{Cio00b}
Cioni, M. R., Habing, H. J., \& Israel, F. P. 2000b,
A\&A, 358, L9

\bibitem{Fea89}
Feast, M. W., Glass, I. S., Whitelock, P. A., \& Catchpole,
R. M. 1989, MNRAS, 241, 375

\bibitem{Fra92}
Franx, M., \& de Zeeuw, P. T. 1992, ApJ, 392, L47

\bibitem{Fre83}
Freeman, K. C., Illingworth, G., \& Oemler, A. 1983, ApJ, 272, 488

\bibitem{Fre87}
Freeman, K. C. 1987, ARA\&A, 25, 603

\bibitem{Gar96}
Gardiner, L. T., \& Noguchi, 1996, MNRAS, 278, 191

\bibitem{Gil89}
Gilmore, G., Wyse, R. F. G., \& Kuijken, K. 1989, 27, 555

\bibitem{Gil90}
Gilmore, G., King, I. R., \& van der Kruit, P. C. 1990, 
The Milky Way as a Galaxy (Mill Valley: University Science Books)

\bibitem{Gou97}
Gould A., Bahcall J. N., \& Flynn C. 1997, ApJ, 482, 913

\bibitem{Hel99}
Helmi, A., White, S. D. M., de Zeeuw, P. T., \& Zhao, H. 1999, Nature, 
402, 53

\bibitem{Iba01}
Ibata, R., Lewis, G. F., Irwin, M., Totten, E., \& Quinn, T. 2001,
ApJ, 551, 294

\bibitem{Joh96}
Johnston, K. V., Hernquist, L., \& Bolte, M. 1996, ApJ, 465, 278

\bibitem{Joh98}
Johnston, K. V., 1998, ApJ, 495, 297

\bibitem{Kui89}
Kuijken K., \& Gilmore G. 1989, MNRAS, 239, 605

\bibitem{Kui91}
Kuijken K., \& Gilmore G. 1991, ApJ, 367, L9

\bibitem{Kui94}
Kuijken K., \& Tremaine, S. D. 1994, ApJ, 421, 178

\bibitem{Lin95}
Lin, D. C. N., Jones, B. F., \& Klemola, A. R. 1995, ApJ, 439, 652

\bibitem{Luk92}
Luks, Th., \& Rohlfs, K. 1992, A\&A, 263, 41

\bibitem{Maj93}
Majewski, S. R. 1993, ARA\&A, 31, 575

\bibitem{Mar98}
Martin, J. C., \& Morrison, H. L. 1998, AJ, 116, 1724

\bibitem{Mae88}
Meatheringham, S. J., Dopita, M. A., Ford, H. C., \& Webster, B. L. 1988,
ApJ, 327, 651

\bibitem{Moo94}
Moore, B., \& Davis, M. 1994, MNRAS, 270, 209

\bibitem{Nik00}
Nikolaev, S., \& Weinberg, M. D. 2000, ApJ, 542, 804

\bibitem{Ode01}
Odenkirchen, M., et al. 2001, ApJ, 548, L165

\bibitem{Oll00}
Olling, R.P., \& Merrifield, M. R. 2000, MNRAS, 311, 361

\bibitem{Pfe94}
Pfenniger, D., Combes, F., \& Martinet, L. 1994, A\&A, 285, 79

\bibitem{Zar95}
Rix, H.-W., \& Zaritsky, D. 1995, ApJ, 447, 82

\bibitem{Sac93}
Sackett, P. D., Gould, A. 1993, ApJ, 419, 648

\bibitem{Sac98} 
Sackett, P. D. 1998, in Galaxy Dynamics, ASP Conf. Series 182,
Merritt, D. R. Valluri, M., \& Sellwood, J. A., eds., p.~393
(San Francisco: ASP)

\bibitem{Sah94}
Sahu, K. C. 1994, Nature, 370, 275

\bibitem{Sch97}
Schoenmakers, R. H. M., Franx, M., \& de Zeeuw, P. T. 1997, MNRAS, 292, 349

\bibitem{Scho92}
Schommer, R. A., Suntzeff, N. B., Olszewski, E. W., \& Harris, H. C.
1992, AJ, 103, 447

\bibitem{Sea99}
Searle, L., \& Zinn, R. 1978, ApJ, 225, 357

\bibitem{vdM91} 
van der Marel, R. P. 1991, MNRAS, 248, 515

\bibitem{vdM01a} 
van der Marel, R. P. 2001, AJ, in press [astro-ph/0105340]

\bibitem{vdM01b} 
van der Marel, R. P., \& Cioni, M. R. 2001, AJ, in press [astro-ph/0105339]

\bibitem{Wei00b}
Weinberg, M. D. 2000, ApJ, 532, 922

\bibitem{Wei00}
Weinberg, M. D., \& Nikolaev, S. 2000, ApJ, 548, 712

\bibitem{Wes97}
Westerlund, B. E. 1997, The Magellanic Clouds (Cambridge: Cambridge
University Press)

\bibitem{Zar97b}
Zaritsky, D., \& Lin, D. N. C. 1997, AJ, 114, 2545

\bibitem{Zar99b}
Zaritsky, D., Shectman, S. A., Thompson, I., Harris, J., \& Lin, D. N. C.
1999, AJ, 117, 2268

\bibitem{Zha00}
Zhao, H. S., \& Evans, N. W. 2000, ApJ, 545, L35

\end{thebibliography}
\end{document}